\documentstyle[aps]{revtex}
\begin{document}
\title{Quantum State Measurement by Realistic Heterodyne Detection}
\author{Matteo G. A. Paris \thanks{E-mail Address: PARIS@PV.INFN.IT.} }
\address{Arbeitsgruppe 'Nichtklassiche Strahlung' der Max-Planck-Gesellschaft
an der Humboldt-Universitat zu Berlin, \\
Rudower Chaussee 5, 12489 Berlin,Germany \thanks{Permanent Address:
Dipartimento di Fisica 'Alessandro Volta', Universit\'a di Pavia, v.
Bassi 6, 27100, Pavia, Italia.}}
\date{4 Aug 1995}
\twocolumn
\maketitle
\begin{abstract}
The determination of the quantum properties of a single mode radiation field
by heterodyne or double homodyne detection is studied. The realistic case of
not fully efficient photodetectors is considered. It is shown that a large
amount of quite {\em precise} information is avalaible whereas the completeness
of such information is also discussed. Some examples are given and the special
case of states expressed as a finite superposition of number states is
considered in some detail.
\end{abstract}
\vspace{5pt}\noindent
{1994 {\bf PACS} number(s): 42.50.Dv; 03.65.Bz; 42.65.Ky.}
\vspace{10pt}\par
\section{Introduction}
In order to get information about a quantum state one has to measure some
observable. A question immediately arises: is this information complete ?
Namely, does it fully specify the quantum state under examination ?
The answer is generally negative: the measurement of only one observable
show up only an aspect of a physical system whereas a complete description
requires a deeper inspection. However, the measurement of several different
observables could require considerable efforts. Therefore it is a matter of
interest to compare the simplicity and the feasibility of a measurement, or
a set of measurements, with the provided amount of information. In addition,
one has to weigh up the precision of such an information.
\par
The complete description of a quantum state is contained in the density
operator $\hat\rho$, or for pure state in the wave function. Therefore, a
measurement leading to the density matrix in some representation provides, in
principle, all the knowable information about the measured state. It has
been shown theoretically \cite{VR} that the Wigner function $W_0
(\alpha,\bar\alpha)$ \cite{Wor} of a field mode can be reconstructed starting
from the homodyne measurements of a continuos set $\varphi \in [-\pi,\pi)$ of
field quadratures $\hat x_{\varphi} =1/2
(ae^{-i\varphi}+a^{\dag}e^{i\varphi})$.
Later this procedure has been applied to coherent and squeezed states
\cite{Raymers} using a finite set of phases $\varphi_i$, $i=1,...27$, upon a
smoothing on experimental data. More recently a procedure suitable to finite
sampling has been suggested for recovering matrix elements in the Fock
representation \cite{tomo} and latest developments have extended the method to
data coming from inefficient detectors \cite{ulf}.
These various procedures,
unitarily referred to as {\em quantum tomography}, provide a nice and
powerful tools
for investigating the quantum properties of radiation field
leading to the maximum information achievable on the measured state. However,
they require the detection of many field quadratures $\hat x_{\varphi_{j}}$,
$j=,1,...,N$ and thus a lot of repeated measurements on the state under
examination. A systematic approach to precision of quantum tomography is in
progress \cite{sterpi} however, a preliminary study \cite{sensi} has indicated
that its determination of some field properties can be very noisy relative to
the direct measurements of the same quantities. \par

\section{Realistic Heterodyne Detection}
In this paper a different approach to quantum state measurement will be
investigated. The crucial remark is that the density matrix in the coherent
state representation $\langle\alpha |\hat\rho |\alpha\rangle$ is a positive
definite function and thus can be directly measured for any quantum state of
radiation field. Indeed, it is known that the so called Husimi $Q$-function
$W_{-1} (\alpha,\bar\alpha) = \langle\alpha |\hat\rho |\alpha\rangle$
represents the outcomes probability distribution
\cite{rapid,altri,ripha}
of both the heterodyne \cite{shap} and the double homodyne \cite{otto}
detection scheme when equipped with ideal photodetectors. On the other hand
its smoothed versions
\begin{equation}
W_{s} (\alpha,\bar\alpha) =\int \frac{d^2\beta}{\pi}\;
W_{-1} (\beta,\bar\beta )\;\exp\{-2\frac{|\alpha -\beta |^2}{1+s}\}
\;,
\label{ssmooth}
\end{equation}
emerge from realistic devices in which not fully efficient detectors are
involved. The parameter $s$ depends on quantum efficiency as
\cite{reali,opti}
\begin{equation}
s=1-2\eta^{-1}
\;.
\end{equation}
\par
Starting from heterodyne, or equivalently from double homodyne \cite{same},
detection we can evaluate some quantity $O$ of interest as an average over
the experimental distribution
\begin{equation}
O=\langle \hat O \rangle = \int \frac{d^2\alpha}{\pi}\;
W_{s} (\alpha,\bar\alpha )\; {\cal F}_s [\hat O] (\alpha,\bar\alpha )
\;,
\label{av1}
\end{equation}
where ${\cal F}_s [\hat O] (\alpha,\bar\alpha )$ is a non operatorial
function related to the $s$-ordering, in the boson operator expansion, of the
corresponding observable $\hat O$ \cite{cahill}.
Denoting by $\{\hat O\}_s$ the $s$-ordered form of an operator we have
for example
\begin{eqnarray}
&a^{\dag}a= \{a^{\dag}a \}_s + \frac{1}{2} (s-1) &\nonumber\\
&a^{\dag 2}a^2 =\{a^{\dag 2}a^2 \}_s +2 (s-1) \{ a^{\dag}a\}_s
+ \frac{1}{2} (s-1)^2 &
\;,
\label{nmean}
\end{eqnarray}
and thus
\begin{eqnarray}
&{\cal F}_s [\hat n] (\alpha,\bar\alpha ) = |\alpha |^2 + \frac{1}{2} (s-1)&
\nonumber\\
&{\cal F}_s [\widehat{n^2 }] (\alpha,\bar\alpha ) = |\alpha |^4
+ (2s-1) |\alpha |^2 + \frac{1}{2} s(s-1) &
\;,
\label{nfluct}
\end{eqnarray}
for the mean photon number and for the number fluctuations.
Similarly, we can investigate the squeezing properties of the examined state
by means of the $s$-ordering of the field quadrature fluctuations
\begin{equation}
{\cal F}_s \left[ \widehat{x^2_{\varphi}} \right]
(\alpha,\bar\alpha ) =  \frac{1}{4}
\left[ \alpha^2 e^{-2i\varphi} + \bar\alpha^2 e^{2i\varphi}
+ 2 | \alpha|^2 +s \right]
\;,
\label{ffluct}
\end{equation}
and also checking the uncertainty product.
This procedure can be generalized in order to evaluate any field correlation
which is of interest. In fact, any $t$-ordered moment
$\{a^{\dag n}a^{n+d} \}_t$, with arbitrary $t$, can be written in terms of
a {\em finite} number of $s$-ordered moment
by means of the formula \cite{cahill}
\FL\begin{eqnarray}
&&\{a^{\dag n}a^{n+d}\}_t=  \nonumber\\
&&\qquad\sum_{k=0}^{n}\frac{(d+n)!}{(d+k)!}
\left(\begin{array}{c} n \\ k \end{array}\right)
\left(\frac{s-t}{2}\right)^{n-k} \!\!\!\!
\{a^{\dag n}a^{n+k}\}_s
\;,
\label{momconn}
\end{eqnarray}
where also $s$ is arbitrary.
The expectation value of any $t$-ordered moment can thus be evaluated
by an average over the different ordered distribution
$W_s (\alpha,\bar\alpha)$. The special
case in which the parameters $t$ and $s$ are choosen
to be $t=1$ and $s=1-2\eta^{-1}$ is
of interest as it allows to obtain the normal ordered field correlations
$\langle a^{\dag n}a^{n+d} \rangle$ in terms of a finite numbers of
heterodyne {\em measured moments} $\langle a^{\dag k}a^{k+d} \rangle_{\eta}$.
We have
\FL\begin{eqnarray}
\langle a^{\dag n}a^{n+d} \rangle &=&\sum_{k=0}^{n} \frac{(d+n)!}{(d+k)!}
\left(\begin{array}{c} n \\ k \end{array}\right)
\left(-\frac{1}{\eta}\right)^{n-k} \nonumber\\
&& \nonumber\\
&\times&\int\! \frac{d^2\alpha}{\pi}\;
W_{1-2\eta^{-1}} (\alpha,\bar\alpha )\; \alpha^{k+d} \bar\alpha^k
\;,
\label{field}
\end{eqnarray}
or in a more compact form
\FL\begin{eqnarray}
\langle a^{\dag n}a^{n+d}\rangle &=&
\frac{(-)^n n!}{\eta^n}
\nonumber\\
&\times&
\int \frac{d^2\alpha}{\pi} \;
W_{1-2\eta^{-1}} (\alpha,\bar\alpha )
\alpha^d \;L_n^d (\eta |\alpha |^2 )
\label{lag}
\end{eqnarray}
where $L_n^d (x)$ denotes Laguerre polynomials.
An interesting application of Eq. (\ref{lag}) lies in the evaluation of the
expectation value $\langle e^{i\hat n\phi}\rangle$ of the shift operator.
Starting from normal ordering
\begin{equation}
e^{i\hat n\phi}=\sum_k (e^{i\phi-1})^k a^{\dag
k}a^k /k!
\;,
\label{expo}
\end{equation}
we have, in fact
\begin{eqnarray}
\langle e^{i\hat n\phi}\rangle = &&
\int \frac{d^2\alpha}{\pi}\; W_{1-2\eta^{-1}} (\alpha,\bar\alpha )
\nonumber\\
&\times& \qquad
\sum_{k=0}^{\infty} \left(\frac{1-e^{i\phi}}{\eta}\right)^k  \!
L_k (\eta |\alpha |^2 )
\;.
\end{eqnarray}
Then, using properties of Laguerre polynomials \cite{grad}, we arrive at the
formula (valid for $0 \leq \phi< \arccos (1-\eta^2/2)$]
\FL\begin{eqnarray}
\langle e^{i\hat n\phi}\rangle =
\frac{\eta}{\eta-1+e^{i\phi}}
&&\int \frac{d^2\alpha}{\pi}\; W_{1-2\eta^{-1}} (\alpha,\bar\alpha )\;
\nonumber\\
&& \nonumber\\
&\times& \exp\{\frac{\eta (1-e^{i\phi})|\alpha |^2}{1-e^{i\phi}-\eta} \}
\label{star2}
\;.
\end{eqnarray}
\par
Eq. (\ref{av1}) is also suitable for a reliable estimation of
errors in the determination of the various expectation values. In
practical situation, in fact, one has at disposal a finite sample of
$W_{s} (\alpha,\bar\alpha )$ and thus the integral in formula (\ref{av1}) has
to replaced by the corresponding statistical sampling
\begin{equation}
\overline{O} = \sum_{j\in \hbox{\scriptsize data}}
W_{s} (\alpha_j,\bar\alpha_j ) {\cal F}_s [\hat O] (\alpha_j,\bar\alpha_j )
\;.
\label{sample}
\end{equation}
According to the law of large numbers $\overline{O}$ approaches
$\langle \hat O\rangle$ in the limit of infinite number of sampling data,
whereas for finite sample we have a confidence interval $\overline{O}\pm
\delta O$, with $\delta O$ given by
\begin{equation}
\delta O = \sqrt{\sum_{j\in \hbox{\scriptsize data}}
W_{s} (\alpha_j,\bar\alpha_j ) {\cal F}^2_s [\hat O] (\alpha_j,\bar\alpha_j )
-{\overline{O}}^2}
\;.
\label{erro}
\end{equation}
Some examples of the present reconstruction procedure can be given by means
of numerical simulations of realistic heterodyne detection. In Fig.
\ref{f:number} I report the results for the mean photon number
$\langle \hat n \rangle$ determination at different values of the quantum
efficiency for coherent states of different amplitudes.
Fig. \ref{f:number}a shows the results from heterodyne detection and
Fig. \ref{f:number}b that ones from a direct photodetection.
The two determinations are also compared in Fig. \ref{f:number}c.
In making such a comparison the same number of repeated measurements on
the field have to be considered. In a scheme of $N$ repeated measurements of
the quantity $x$, the accuracy $\delta x$ rescales as $\delta x \propto
N^{-1/2}$. The proportionality costant generally depends on the kind of
detection. If the outcomes $\bar x$ are distributed around the true value
$x$ according to the probability $p(\bar x | x)$, the error for $N$ repeated
measurements is always bounded by the Cramer-Rao inequality \cite{cramer}
$\delta x \geq (FN)^{-1/2}$, F being the Fisher information
$F= \int d\bar x \; [\partial_x p(\bar x | x)]^2 /p(\bar x | x) $. For Gaussian
distributed data one has $F= 1/\sigma^2$, with $\sigma^2$ the variance of the
distribution, and the lower bound for precision is achieved.
In practical situations, in order to evaluate the precision $\delta x$, one
takes advantage of the central limit theorem \cite{cramer},
which assures that the partial
averages over a block of $N_b$ data is always Gaussian distributed around the
global average over many blocks. Thus, one evaluates precision by dividing the
ensemble of data into subensembles, and then calculates the r.m.s. deviation
of subensemble averages with respect to the global one. \par
{}From Fig. \ref{f:number} it is apparent that the method works also
for low efficiency of the
photodetectors even though the results are slightly more noisy than ones from
direct detection. However, this level of introduced noise can be considered as
admissible in sight of the further information available from the same data
sample. Moreover, it has to be noticed (see Ref. \cite{sensi}) that
tomographic determination of $\langle \hat n \rangle$ introduces a very large
amount of noise, even for unit quantum efficiency. Fig. \ref{f:field}
illustrates the application in recovering field fluctuations
on a squeezed state and a number state
for different values of the quantum efficiency.
\par
About the determination of the phase some further considerations are in order.
Heterodyne detection, in fact, is by itself a phase detectors as the marginal
distribution
\begin{equation}
P_s (\phi) = \int_0^{\infty} \rho\;d\rho \; W_s (\rho e^{i\phi},\rho e^{-i\phi}
)
\;,
\label{sphase}
\end{equation}
represents a phase distribution of the measured state \cite{opti,14}.
We have thus at disposal not only the mean value of the phase and the
fluctuations but also the whole probability distribution. The
distribution in Eq. (\ref{sphase}) does not coincide (even for $\eta=1$ ) with
the canonical phase distribution \cite{ripha,14}
\begin{equation}
P(\phi) = \langle e^{i\phi}|\hat\rho |e^{i\phi}\rangle =\frac{1}{2\pi}
\sum_{n,m}^{\infty} e^{i(n-m)\phi} \rho_{n,m}
\;,
\label{london}
\end{equation}
and it is generally broadened relative to (\ref{london}).
In Fig. \ref{f:phase} the phase distribution obtained for a squeezed vacuum is
reported for experiments carried out with different values of the quantum
efficiency. The distributions broaden when the quantum efficiency decreases
but the crucial information about phase bifurcation \cite{varro} is not lost
also for for very inefficient detectors. \par
The results obtained until now can be summarized in the following assertions:
i) starting from heterodyne detection many properties of the measured state
can be determined at one go; ii) this determination is slightly more noisy
than the corresponding ones from direct measurements of the same quantities
one at times, even for unit quantum efficiency of the photodetectors. However,
this behaviour is not unexpected as heterodyne detection involves the joint
measurement of non commuting observables, and thus the unavoidably addition of
noise by first principles \cite{yuen82,good}.
This is not the case of quantum tomography where each homodyne measurement is
independently performed and noise is introduced by data processing itself.
\par
\section{Density Matrix Reconstruction}
Let us now deal with the completeness of the information coming from
heterodyne detection. Is it possible, as an example, to determine the whole
number distribution ? The matrix elements $\rho_{n+k,n}$ in the Fock
representation are the expectation values of the generalized projectors
\begin{equation}
\hat P_{n,n+k}=|n\rangle\langle n+k |= \frac{a^{\dag n}}{\sqrt{n!}}\;
|0\rangle\langle 0|\; \frac{a^{n+d}}{\sqrt{(n+d)!}}
\;.
\label{proje}
\end{equation}
Using the Louisell expansion of the
vacuum \cite{louisell}
\begin{equation}
|0\rangle\langle 0| = \lim_{\varepsilon \rightarrow 1^-} \sum_{p}
\frac{(-\varepsilon )^p}{p!}  a^{\dag p} a^{p}
\label{luisell}\;,
\end{equation}
we can write $\hat P_{n,n+k}$ in terms of normal ordered moments
\begin{equation}
\hat P_{n,n+k} = \frac{1}{\sqrt{n!(n+k)!}}
\lim_{\varepsilon \rightarrow 1^-}
\sum_{p} \frac{(-\varepsilon )^p}{p!}  a^{\dag n+p} a^{n+p+k}
\;. \label{ord}
\end{equation}
Eq. (\ref{ord}) is suitable to apply Eq. (\ref{momconn}). After a
straighforward
calculation we get the reconstruction formula (\ref{av1}) for the matrix
elements
\begin{eqnarray}
\rho_{n+k,n} &=& (-)^n \sqrt{\frac{n!}{(n+k)!}} \sum_{q=n}^{\infty}
\left(\frac{1}{\eta}\right)^q
\left(\begin{array}{c} q\\n \end{array}\right) \nonumber\\
&\times&
\int\!\frac{d^2\alpha}{\pi}\;W_{1-2\eta^{-1}} (\alpha,\bar\alpha )\;
\alpha^k L_q^k (\eta |\alpha |^2)
\;.
\label{rho}
\end{eqnarray}
Unfortunately, Eq. (\ref{rho}) is not suitable for statistical sampling as
the interchange of integration and summation is not
mathematically allowed \cite{baltin}. The analytical expression of
$W_{s} (\alpha,\bar\alpha )$ is needed and thus some smoothing procedure on
sampled data is required, unavoidably introducing some {\em a priori}
hypothesis on the state under examination \cite{failure}. However,
Eq. (\ref{rho}) is far from being
a purely formal tool. In many situations, in fact, one has some information
about the considered state and thus some suggestions on parametryzing
Wigner functions. Moreover, the distributions  $W_{s} (\alpha,\bar\alpha )$
for $s \leq -1$ are smoothed functions by themselves \cite{cahill} and
generally do not exhibit large or fast oscillations. Therefore we may expect
the smoothing not leading to a dramatic lost of information and, at the same
time, to not introduce fake information. The completeness of information
coming from heterodyne detection cannot, anyhow, be claimed in a general way.
\par
The reconstruction of the entire density matrix (in the Fock representation)
and thus a complete description of the state is possible for the special case
of states with a finite number of moments different from zero. Examples of
such a states are provided by finite superpositions of number states
\begin{equation}
|\psi \rangle = \sum_{n=0}^N \psi_n |n\rangle
\;.
\label{super}
\end{equation}
The latter can be produced in different manner in a high-Q cavities
\cite{uno,due} and also by a special non linear interaction \cite{bielo}.
If the moments $a^{\dag n}a^m$ vanish for $n$ or $m$ beyond a certain
value the series in Eq. (\ref{ord}) are actually truncated and Eq. (\ref{rho})
becomes  suitable to a statistical sampling
\FL\begin{eqnarray}
\rho_{n+k,n}&=&\frac{(-)^n}{\eta^n}\sqrt{\frac{n!}{(n+k)!}}
\int \frac{d^2\alpha}{\pi}\;W_{1-2\eta^{-1}} (\alpha,\bar\alpha )\; \alpha^k
\nonumber\\
&& \nonumber\\
&\times&
\sum_{p=0}^{N-n-k}
\left(\begin{array}{c} p+n \\ p \end{array}\right)
\left(\frac{1}{\eta}\right)^p
L_{p+n}^k (\eta |\alpha |^2 )
\;.
\label{rhon}
\end{eqnarray}
The value of $N$ has to be choosen large enough to
ensure the cancellation of any
moment $a^{\dag N+j}a^{N+i}$, $i,j=0,1,...$. In practice one can start
with a large value of $N$ and then optimizing it by means of some stability
criterion. In any case the precise value of $N$ is not needed by the
algorithm. In Table \ref{t:super} I report the results of this procedure
when applied to the state
\begin{equation}
|\psi\rangle = \frac{1}{\sqrt{2}}\left( |0\rangle + i |2\rangle \right)
\;,
\label{0i2}
\end{equation}
using photodetectors with an overall quantum efficiency equal to $\eta=0.9$.
The reliability of the method is apparent. The same degree of precision can
be obtained with lower efficiency using a larger sample.
\par
The problem of reconstructing the density matrix of field-states with finite
numbers occupation has been considered also by other authors, in particular in
the context of the so-called Pauli's phase retrieval problem, where two
experimentally determined probability distributions are needed.
Orlowsky and Paul proposed in \cite{orlowsky} an algorithm to recover the
entire wavefunction (\ref{super}), assuming that the moduli of the
wavefunction are known in the position and momentum representation.
Their method involves solving blocks of nonlinear equations after a
decomposition of the wavefunction into Hermite polynomials. The resulting
phase retrieval scheme is transparent, however it is recursive from
above, namely it determines the coefficients $\psi_n$ from the highest
index $N$ to the lowest.
In addition, the value $N$ of the nonzero components of the
wavefunction has to be known in advance. On the contrary, Eq. (\ref{rhon})
allows recovering of the matrix elements $\rho_{n,m}$ {\it one by one} as an
average over the experimental distribution and the value of $N$ is not needed
by the algorithm.
Bialynicka-Birula and Bialynicki-Birula in \cite{bb} considered the
reconstruction problem starting from the knowledge of the photon number and
phase (Pegg-Barnett) distributions. They suggested two different
algorithms based on recursive iterations of Fast Fourier Transform from the
number representation to the phase domain. Their algorithms work only for pure
states whereas the present one can also be applied in the general case.
In fact, the only requirement for writing Eq. (\ref{rho}) in the
sampling-suited form (\ref{rhon}) is that the measured state possesses only a
finite number of moments different from zero. This condition can obviously be
fulfilled also by finite mixtures. It has also to be mentioned that a
detection scheme for the Pegg-Barnett phase distribution has not been devised
yet. Thus the phase distribution needed by the algorithms in \cite{bb} can
only be inferred by other state measurement schemes such as homodyning or
quantum tomography. \par
Apart from the above considerations I consider the reliability of the present
method in evaluating the confidence interval on matrix elements determinations
as its main advantage.
\par
\section{Conclusion}
In conclusion, quantum state measurement by heterodyne or double homodyne
detectors has been shown to provide a large amount of quite precise
information. It cannot lead to a complete specification of the measured state
due to the singularity in some $s$-ordering ($s \leq -1$) of operators.
To the knowledge of the author it represents, at current time, the best
compromise between the conflicting necessity of precise and complete
state measurement.
\par
I would thank 'Angelo Della Riccia' foundation for a research
grant and Prof. Harry Paul for valuable hints. I am also very grateful to
Valentina De Renzi for crucial encouragements.

\begin{figure}
\caption{Mean photon number determination by a simulated heterodyne and direct
detections for different values of the quantum efficiency $\eta$. In (a) the
results for three different coherent states of amplitude
$\alpha =1$ (circle), $\alpha =2$ (square) and $\alpha =3$ (triangle) are
reported for heterodyne detections. In (b) are reported the results from
direct detections with the same number of repeated measurements on the same
coherent states. The confidence intervals of both the determinations are
evaluated as usual, by dividing the whole sample of $10^5$ data in subensembles
and then calculating r.m.s. deviation with respect to the global average
(see text). In (c) the accuracy of the two kinds of determination is compared.
The noise (in dB) added by heterodyne detection is shown as the ratio
between the corresponding confidence intervals. }
\label{f:number}
\end{figure}

\begin{figure}
\caption{Simulated determination of field fluctuations
$\overline{\Delta x^2_{\varphi}}$  for $\varphi=0,\pi/2$.
Results for a squeezed state of total energy $\langle\hat n\rangle=1$
equally distributed between coherent amplitude and squeezing and a number
state $\hat\rho =|1\rangle\langle 1|$ are reported.
The upper and the lower curves are for the squeezed state ($\varphi=0,\pi/2$
respectively), the central one the result for number state (isotropic field
distribution). Samples of $10^5$ data are used and the confidence intervals are
evaluated as in Fig.
\protect\ref{f:number}.}
\label{f:field}
\end{figure}

\begin{figure}
\caption{Phase distribution from heterodyne detection for a squeezed vacuum
with squeezing photons $\langle \hat n\rangle =1$. The distributions are
obtained with a sample of $10^5$ data for $\eta=0.25,0.5,0.75,1.0$.
Broader distributions correspond to lower values of $\eta$.}
\label{f:phase}
\end{figure}
\newpage
\widetext
\begin{table}
\caption{Reconstructed density matrix along with the confidence intervals for
the state $|\psi\rangle = 1/ \protect\sqrt{2} ( |0\rangle + i |2\rangle )$.
The experiment has been simulated for quantum efficiency $\eta=0.9$ using a
sample of $10^6$ data.}
{\small
\begin{tabular}{cccc}
   $.502\pm .024$ & $(.004-i.003)\pm(.022+i.021)$ &
   $(.001-i.493)\pm(.037+i.038)$                  & $\cdots$ \\
 & & & \\
   $(.004+i.002)\pm(.021+i.022)$ & $-.003\pm.053$ &
   $(-.003+i.002)\pm(.018+i.018)$                 & $\cdots$ \\
 & & & \\
   $(.001+i.493)\pm(.037-i.038)$ & $(-.003-i.002)\pm(.018+i.018)$ &
   $.500\pm .031$                                 & $\cdots$ \\
 & & & \\
   $\cdots$&$\cdots$&$\cdots$&$ -.004\pm.063$
\end{tabular}   }
\label{t:super}
\end{table}
\narrowtext

\begin{references}
\bibitem{Wor} I adopt the usual definition for the generalized Wigner
functions
\begin{equation}
W_s (\alpha,\bar\alpha ) = \int \frac{d^2\lambda}{\pi} \hbox{Tr}
\left\{
e^{\lambda a^{\dag} +\bar\lambda a+\frac{1}{2}s|\lambda |^2}
\right\}
e^{\lambda \bar\alpha +\bar\lambda \alpha}
\nonumber\;,
\end{equation}
in which the values $s=1, 0, -1$ respectively correspond to Glauber $P$,
symmetric Wigner and Husimi $Q$ functions.
\bibitem{same} Throughout the paper any reference to the heterodyne detectors
is also valid for the double homodyne detectors.
\bibitem{failure} The needing of smoothing lies on the singularity
of the antinormal ordering. The functions ${\cal F}_s
[\hat P_{n,n+k}] (\alpha,\bar\alpha )$ have been obtained as a series
of well-behaved functions however, actually they can be written in terms of
the Dirac-delta distribution in the complex plane and its derivatives
$$ {\cal F}_s [\hat P_{n,n+k}] (\alpha, \bar\alpha ) =
\frac{1}{\sqrt{n!(n+k)!}} e^{|\alpha |^2} \frac{\partial^{2n+k}}{\partial
\alpha^{n+d}\partial\bar\alpha^{n}} \delta (\alpha )
\nonumber.$$
\end{references}
\end{document}